\documentclass[fleqn,usenatbib,useAMS]{mnras}
 \usepackage{graphicx}
 \usepackage{amsmath}
 \usepackage{amssymb}
 \usepackage[T1]{fontenc}
 \usepackage{ae,aecompl}
 \usepackage{txfonts}
 \usepackage{enumerate}

 \title[Orbital evolution of asteroid 2020~AV$_{2}$]
       {On the orbital evolution of 2020~AV$_\mathbf{2}$, 
        the first asteroid ever observed to go around the Sun inside the orbit of Venus} 

 \author[C. de la Fuente Marcos and R. de la Fuente Marcos]
        {C.~de~la~Fuente~Marcos$^{1}$\thanks{E-mail: nbplanet@ucm.es}
         and
         R. de la Fuente Marcos$^{2}$ \\
         $^1$Universidad Complutense de Madrid,
             Ciudad Universitaria, E-28040 Madrid, Spain \\
         $^2$AEGORA Research Group,
             Facultad de Ciencias Matem\'aticas,
             Universidad Complutense de Madrid,
             Ciudad Universitaria, E-28040 Madrid, Spain}
 \date{Accepted 2020 February 7.
       Received 2020 February 3;
       in original form 2020 January 22}
 \pubyear{2020}
 
\begin{document}
  \label{firstpage}
  \pagerange{\pageref{firstpage}--\pageref{lastpage}}
  \maketitle

  \begin{abstract}
     The innermost section of the Solar system has not been extensively 
     studied because minor bodies moving inside Earth's orbit tend to 
     spend most of their sidereal orbital periods at very low solar 
     elongation, well away from the areas more frequently observed by programs 
     searching for near-Earth objects. The survey carried out from the Zwicky 
     Transient Facility (ZTF) is the first one that has been able to detect 
     multiple asteroids well detached from the direct gravitational 
     perturbation of the Earth--Moon system. ZTF discoveries include 
     2019~AQ$_{3}$ and 2019~LF$_{6}$, two Atiras with the shortest periods 
     among known asteroids. Here, we perform an assessment of the orbital 
     evolution of 2020~AV$_{2}$, an Atira found by ZTF with a similarly short 
     period but following a path contained entirely within the orbit of Venus. 
     This property makes it the first known member of the elusive Vatira 
     population. Genuine Vatiras, those long-term dynamically stable, are 
     thought to be subjected to the so-called von~Zeipel--Lidov--Kozai 
     oscillation that protects them against close encounters with both Mercury 
     and Venus. However, 2020~AV$_{2}$ appears to be a former Atira that 
     entered the Vatira orbital domain relatively recently. It displays an 
     anticoupled oscillation of the values of eccentricity and inclination, 
     but the value of the argument of perihelion may circulate. Simulations 
     show that 2020~AV$_{2}$ might reach a 3:2~resonant orbit with Venus in 
     the future, activating the von~Zeipel--Lidov--Kozai mechanism, which in 
     turn opens the possibility to the existence of a long-term stable 
     population of Vatiras trapped in this configuration.   
  \end{abstract}


  \section{Introduction}
     Any minor body following an orbit contained entirely within the orbit of Earth is part of a distinctive dynamical class known as Atiras
     or Interior Earth Objects (IEOs) that currently has 21 known members. Such objects have aphelion distances, $Q$, $<$0.983~au and cannot 
     experience close flybys with the Earth--Moon system. Similarly, objects with $Q$$<$0.718~au could be called Interior Venus Objects 
     (IVOs) although there is not yet a consensus among the scientific community regarding the name of this dynamical class. Such objects 
     can experience close flybys with neither the Earth--Moon system nor Venus, leaving Mercury as their sole direct perturber. 
     \citet{2012Icar..217..355G} argued that minor bodies with 0.718~au$<Q<$ 0.983~au should be called Atiras, after the first named member 
     of this dynamical class, and those with $Q$ in the range 0.307--0.718~au should be known as Venus Atiras or Vatiras. A systematic 
     search for objects with $Q$$<$0.718~au requires a well-designed space mission because they are nearly permanently confined inside the 
     Sun's glare, but discoveries from the ground during favourable visibility windows are also possible.

     The quest for finding the first object with an orbit completely interior to that of Venus started with the Near-Earth Object 
     Surveillance Satellite (NEOSSat) mission \citep{2012LPICo1667.6463H} that has been in operation since 2013, but it has also been 
     attempted from the ground by the Zwicky Transient Facility (ZTF) observing system \citep{2014SPIE.9147E..79S,2017NatAs...1E..71B,
     2020AJ....159...70Y} at Palomar Mountain and EURONEAR \citep{2018A&A...609A.105V} from La Palma. This quest has ended recently with the 
     discovery by ZTF of the first IVO or Vatira, 2020~AV$_{2}$ \citep{2020MPEC....A...99B}. Here, we perform an assessment of the orbital 
     evolution of 2020~AV$_{2}$ using the available data and $N$-body simulations. We explore the possible resonant status of this object, 
     as well as the role of the so-called von~Zeipel--Lidov--Kozai oscillation \citep{1910AN....183..345V,1962P&SS....9..719L,
     1962AJ.....67..591K,2019MEEP....7....1I} on its past and future dynamical evolution. This Letter is organized as follows. In Section~2, 
     we comment on context, data, and methods. The orbital evolution of 2020~AV$_{2}$ is explored in Section~3. Our results are discussed in 
     Section~4 and our conclusions are summarized in Section~5.

  \section{Context, data, and methods}
     \citet{2011epsc.conf..284N} published the first computations that led to the identification of Vatiras numerically and concluded that 
     most objects in this class may exhibit von~Zeipel--Lidov--Kozai oscillations. The von~Zeipel--Lidov--Kozai mechanism requires the 
     concurrent oscillation of the values of eccentricity, $e$, inclination, $i$, and argument of perihelion, $\omega$, in the frame of 
     reference of the invariable plane of the Solar system (see e.g. \citealt{1999ssd..book.....M}). Most Atiras display an anticoupled 
     oscillation of the values of $e$ and $i$, but no libration of the value of $\omega$ \citep{2018RNAAS...2b..46D,2019MNRAS.487.2742D,
     2019RNAAS...3..106D}. Therefore, no known Atiras are subjected to von~Zeipel--Lidov--Kozai oscillations, but stable Vatiras may 
     experience these oscillations \citep{2019MNRAS.487.2742D,2019RNAAS...3..106D}.

     The discovery of 2019~AQ$_{3}$ and 2019~LF$_{6}$ ---two Atiras with the shortest periods among known asteroids--- suggested that 
     Vatiras ought to exist and that they may be numerous because there are dynamical pathways that may turn Atiras into Vatiras and vice 
     versa \citep{2019MNRAS.487.2742D}. Asteroid 2020~AV$_{2}$ was first observed by ZTF on 2019 January 4 \citep{2020MPEC....A...99B}. Its
     newest orbit determination (see Table~\ref{elements}) is based on 135 observations for a data-arc span of 19~d. The value of its 
     aphelion distance, $Q$=0.65377$\pm$0.00012~au, which is the shortest known after that of Mercury, confirms that it is an IVO or Vatira 
     (see above). Asteroid 2020~AV$_{2}$ is relatively large with an absolute magnitude of 16.4~mag (assumed $G=0.15$), which suggests a 
     diameter in the range $\sim$1--8~km for an assumed albedo in the range 0.60--0.01. The value of its semimajor axis, $a$, is 
     0.55542$\pm$0.00010~au, which is very similar to those of 2019~AQ$_{3}$ (0.58866153$\pm$0.00000008~au) and 2019~LF$_{6}$ 
     (0.5553$\pm$0.0002~au). The orbital periods of 2019~LF$_{6}$ and 2020~AV$_{2}$ are virtually the same. De la Fuente Marcos \& de la 
     Fuente Marcos \citeyearpar{2019RNAAS...3..106D} have shown that 2019~LF$_{6}$ is not subjected to the von~Zeipel--Lidov--Kozai 
     oscillation and it is not currently in mean-motion resonance with any planet. However, it is in near-mean-motion resonance with Mercury 
     (12:7), Venus (2:3), Earth (12:29), Mars (2:9), and Jupiter (1:29), in other words, the relevant critical angles do not librate over 
     time about constant values (see e.g. \citealt{2019MNRAS.483L..37D}). Out of these near-mean-motion resonances of 2019~LF$_{6}$, the 
     ones with Earth, Jupiter, and Mars (in this order) are the closest. For both 2019~AQ$_{3}$ and 2019~LF$_{6}$ (and several other known 
     Atiras), the Earth--Moon system and Jupiter are the main secular perturbers \citep{2019MNRAS.487.2742D}. 

     Data in Table~\ref{elements} as well as most input data used in our calculations have been obtained from Jet Propulsion Laboratory's 
     Solar System Dynamics Group Small-Body Database (JPL's SSDG SBDB, \citealt{2015IAUGA..2256293G})\footnote{\url{https://ssd.jpl.nasa.gov/sbdb.cgi}}
     and JPL's \textsc{horizons}\footnote{\url{https://ssd.jpl.nasa.gov/?horizons}} ephemeris system \citep{GY99}. Full details of the
     calculations presented here are discussed in \citet{2012MNRAS.427..728D} and \citet{2019MNRAS.487.2742D}.

%
%----------------------------------------------------------------------------------------------------------------------------------- TABLE I
%------------------------------------------------------------------------------------------------------- Orbital elements asteroids 2020 AV2
%
     \begin{table}
      \centering
      \fontsize{8}{11pt}\selectfont
      \tabcolsep 0.15truecm
      \caption{Values of the Heliocentric Keplerian orbital elements of 2020~AV$_{2}$ and their associated 1$\sigma$ uncertainties. The 
               orbit determination is referred to epoch JD 2459000.5 (2020-May-31.0) TDB (Barycentric Dynamical Time, J2000.0 ecliptic and 
               equinox). Source: JPL's SBDB (solution date, 2020-Feb-01 05:47:18).
              }
      \begin{tabular}{ccccc}
       \hline
        Orbital parameter                                 &   & value$\pm$1$\sigma$ uncertainty \\
       \hline
        Semimajor axis, $a$ (au)                          & = &   0.55542$\pm$0.00010           \\
        Eccentricity, $e$                                 & = &   0.1771$\pm$0.0003             \\
        Inclination, $i$ (\degr)                          & = &  15.872$\pm$0.008               \\
        Longitude of the ascending node, $\Omega$ (\degr) & = &   6.707$\pm$0.004               \\
        Argument of perihelion, $\omega$ (\degr)          & = & 187.31$\pm$0.02                 \\
        Mean anomaly, $M$ (\degr)                         & = & 222.49$\pm$0.09                 \\
        Perihelion, $q$ (au)                              & = &   0.4571$\pm$0.0002             \\
        Aphelion, $Q$ (au)                                & = &   0.65377$\pm$0.00012           \\
        Absolute magnitude, $H$ (mag)                     & = &  16.4$\pm$0.8                   \\
       \hline
      \end{tabular}
      \label{elements}
     \end{table}
%
%-------------------------------------------------------------------------------------------------------------------------------------------
%

  \section{Orbital evolution}
\begin{figure}
       \centering
        \includegraphics[width=\linewidth]{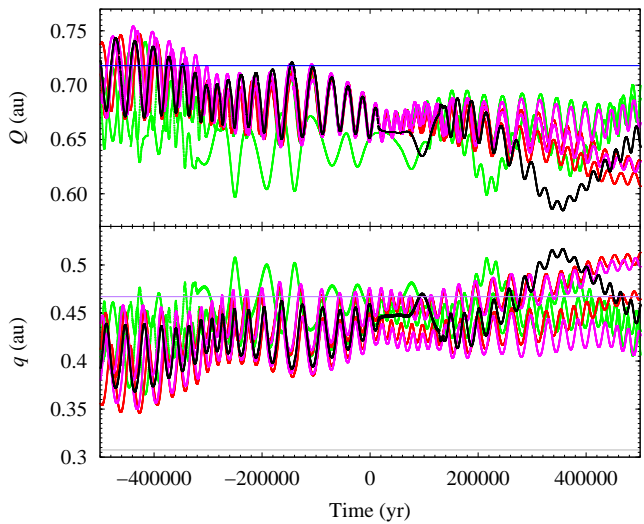}
        \caption{Evolution of the values of the aphelion ($Q$, top panel) and perihelion ($q$, bottom panel) distances of the nominal orbit 
                 (in black) of 2020~AV$_{2}$ as in Table~\ref{elements} and those of control orbits with Cartesian vectors separated 
                 $\pm$3$\sigma$ (in green), $\pm$6$\sigma$ (in red), and $\pm$9$\sigma$ (in magenta) from the nominal values in 
                 Table~\ref{vector}. The perihelion distance of Venus is shown in blue (top panel), the aphelion distance of Mercury is 
                 displayed in purple and its perihelion distance in grey (bottom panel). The nominal orbit evolves towards capture in the 
                 3:2 mean-motion resonance with Venus at 0.552~au.
                }
        \label{Qq}
     \end{figure}

     When considering an orbit determination like the one in Table~\ref{elements}, the first questions that one may want to ask are how 
     reliable is the Vatira dynamical status of 2020~AV$_{2}$ and if it has been a Vatira for a long period of time and if it will remain as 
     such long into the future. Figure~\ref{Qq} shows the evolution of the values of $Q$ and $q$ for the nominal orbit in 
     Table~\ref{elements} in black. We observe a periodic oscillation of the values of $Q$ and $q$ associated with the libration of the 
     value of $e$ as found by \citet{2018RNAAS...2b..46D,2019MNRAS.487.2742D,2019RNAAS...3..106D} for many Atiras. When integrating into the 
     past, the value of $Q$ increases beyond the orbit of Venus, 0.72~au; in other words, the evolution of the nominal orbit suggests that 
     2020~AV$_{2}$ may be a former Atira that was driven into the Vatira orbital realm as a result of multiple relatively distant encounters 
     (i.e. beyond the Hill radius) with Mercury (Hill radius, 0.0012~au) and Venus (0.0067~au). Table~\ref{vector} shows the nominal 
     Cartesian state vector of 2020~AV$_{2}$  and associated uncertainties for the epoch used in Table~\ref{elements}. This information can 
     be used to answer the questions posed above.
%
%---------------------------------------------------------------------------------------------------------------------------------- TABLE II
%------------------------------------------------------------------------------------------------ Geometric Cartesian state vectors 2020 AV2
%
     \begin{table}
      \centering
      \fontsize{8}{11pt}\selectfont
      \tabcolsep 0.15truecm
      \caption{Cartesian state vector of 2020~AV$_{2}$: components and associated 1$\sigma$ uncertainties. Epoch as in Table~\ref{elements}. 
               Source: JPL's SBDB.
              }
      \begin{tabular}{ccccc}
       \hline
        Component                         &   &    value$\pm$1$\sigma$ uncertainty                                \\
       \hline
        $X$ (au)                          & = &    4.468095806605945$\times10^{-1}$$\pm$5.39490272$\times10^{-4}$ \\
        $Y$ (au)                          & = &    4.394371500750955$\times10^{-1}$$\pm$4.11440121$\times10^{-4}$ \\
        $Z$ (au)                          & = &    1.071593681047472$\times10^{-1}$$\pm$1.86856646$\times10^{-4}$ \\
        $V_X$ (au/d)                      & = & $-$1.546838589354595$\times10^{-2}$$\pm$1.81479293$\times10^{-5}$ \\
        $V_Y$ (au/d)                      & = &    1.206070911767420$\times10^{-2}$$\pm$2.18304244$\times10^{-5}$ \\
        $V_Z$ (au/d)                      & = &    3.920491267033128$\times10^{-3}$$\pm$3.58100590$\times10^{-6}$ \\
       \hline
      \end{tabular}
      \label{vector}
     \end{table}
%
%-------------------------------------------------------------------------------------------------------------------------------------------
%

     In addition to that of the nominal orbit determination, Fig.~\ref{Qq} shows the evolution of $Q$ and $q$ for representative control 
     orbits with Cartesian vectors separated $\pm$3$\sigma$ (in green), $\pm$6$\sigma$ (in red), and $\pm$9$\sigma$ (in magenta) from the 
     nominal values in Table~\ref{vector}. They show that 2020~AV$_{2}$ is a statistically robust present-day Vatira and all the orbits 
     compatible with the observations are consistent with the Vatira dynamical type, both thousands of years into the past and the future. 
     Our calculations show that 2020~AV$_{2}$ became detached from Venus about 10$^{5}$~yr ago (Fig.~\ref{Qq}, top panel), but also that its 
     future orbital evolution may lead to being detached from Mercury as well, in about 2$\times$10$^{5}$~yr (Fig.~\ref{Qq}, bottom panel). 
     The unusual evolution into the future of the nominal orbit will be discussed later, but it is the result of capture in the 3:2 
     mean-motion resonance with Venus at 0.552~au.
%
%-------------------------------------------------------------------------------------------------------------------------------------------
%
%
%-------------------------------------------------------------------------------------------------------------------------------------------
%

\begin{figure*}
       \centering
        \includegraphics[width=0.33\linewidth]{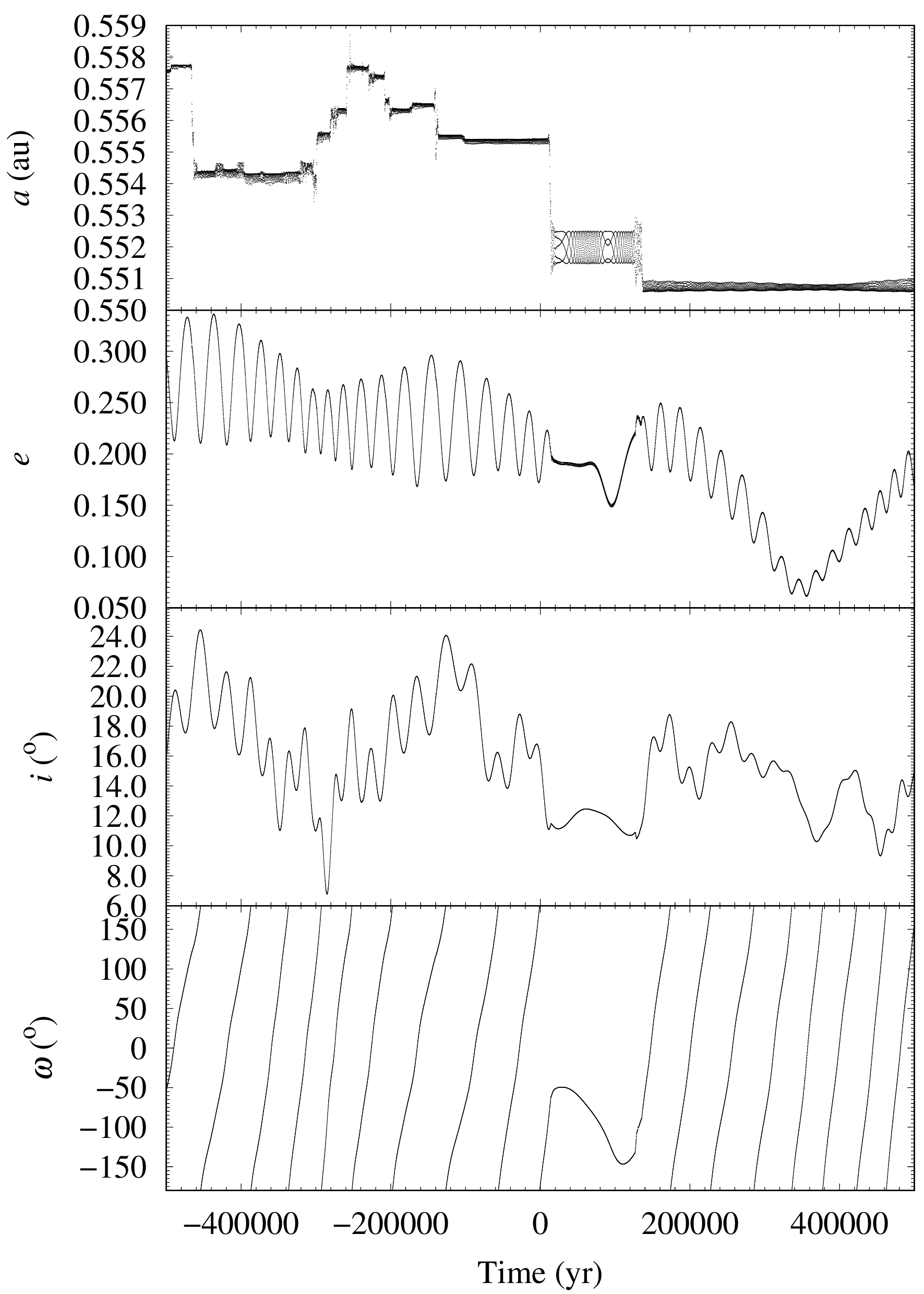}
        \includegraphics[width=0.33\linewidth]{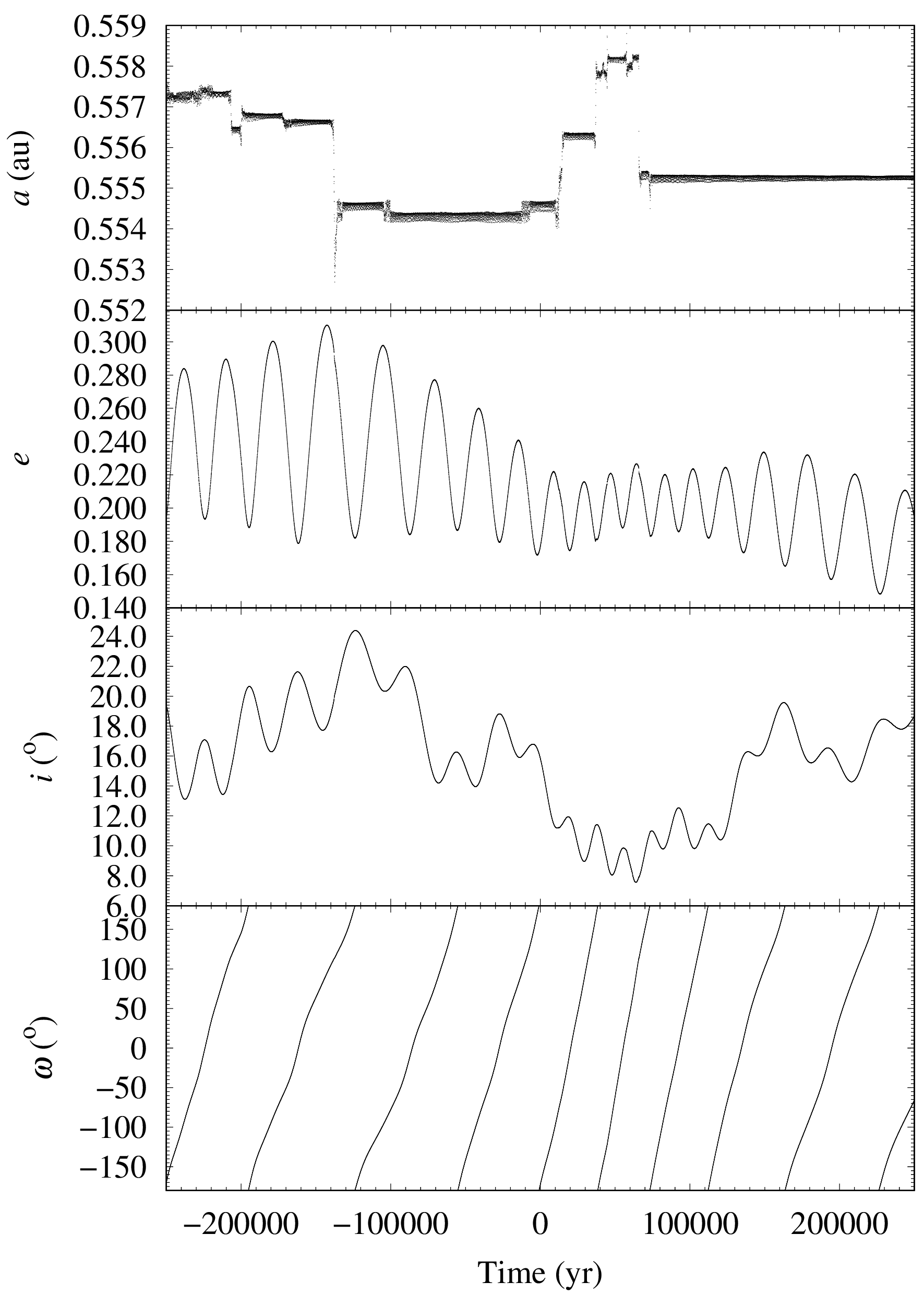}
        \includegraphics[width=0.33\linewidth]{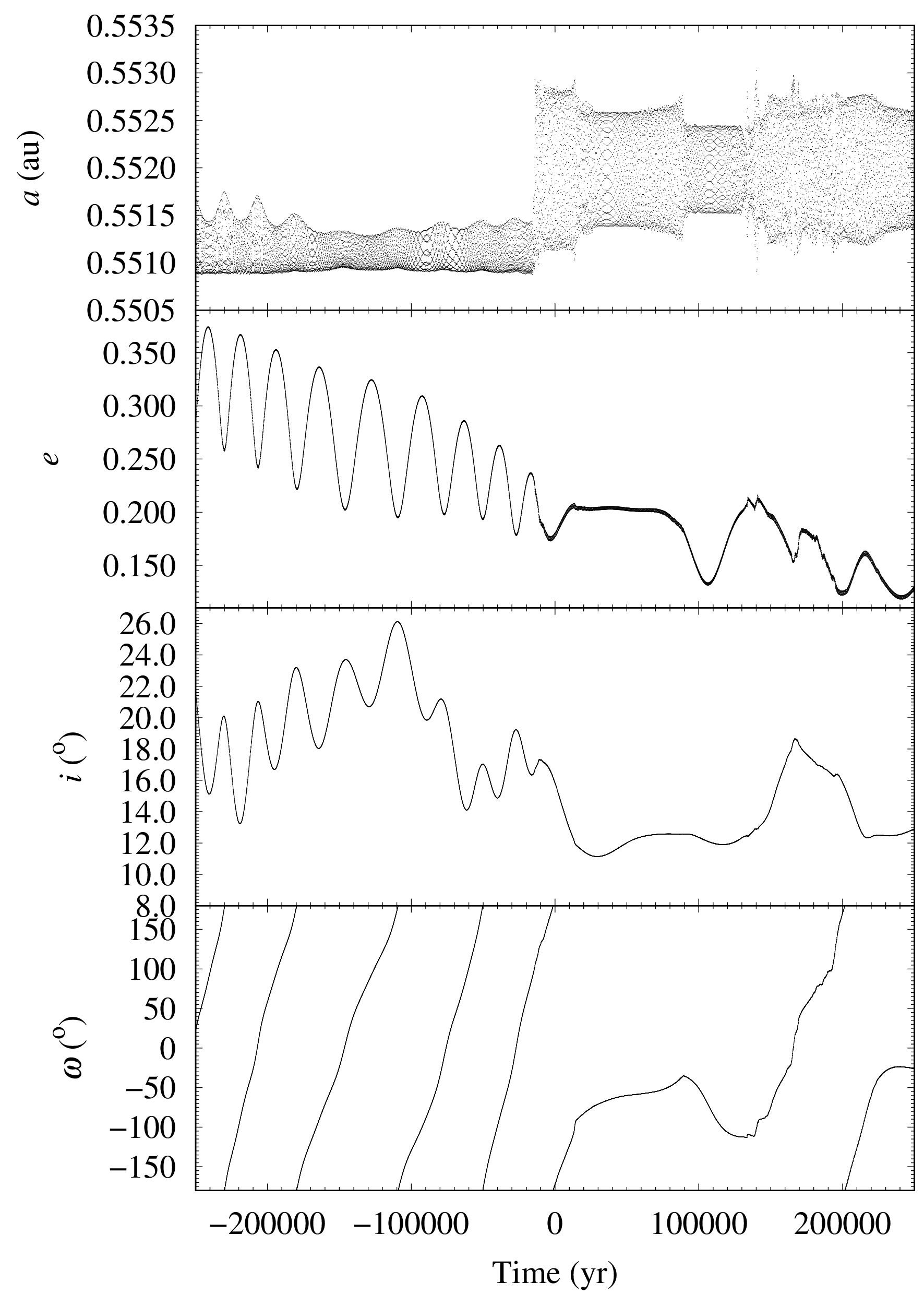}
        \caption{Evolution of the values of the semimajor axis ($a$, top panel), eccentricity ($e$, second to top panel), inclination ($i$, 
                 second to bottom panel), and argument of perihelion ($\omega$, bottom panel) for the nominal orbit of 2020~AV$_{2}$ 
                 (left-hand side set of panels), one close to the $-$6$\sigma$ orbit (central set of panels), and  one close to the 
                 $-$12$\sigma$ orbit (right-hand side set of panels). These values have been computed with respect to the invariable plane 
                 of the system. 
                }
        \label{inv}
     \end{figure*}

     Ngo et al. \citeyearpar{2011epsc.conf..284N} argued that most high-inclination Vatiras exhibit von~Zeipel--Lidov--Kozai oscillations 
     and also that they may have an origin in the main asteroid belt (see their fig.~1). De la Fuente Marcos \& de la Fuente Marcos 
     \citeyearpar{2019MNRAS.487.2742D} showed that Vatiras subjected to von~Zeipel--Lidov--Kozai oscillations may be long-term stable and 
     also that Atens may cross into the Atira orbital domain to become Vatiras at a later time. From our previous analysis, 2020~AV$_{2}$ 
     may have come from near-Earth space and possibly from beyond. Figure~~\ref{inv} shows the evolution of $a$, $e$, $i$ and $\omega$ for 
     the nominal orbit (left-hand side set of panels), and those of control orbits with initial conditions close to $-$6$\sigma$ (central 
     set of panels) and $-$12$\sigma$ values (right-hand side set of panels). Orbits often display the usual anticoupled oscillation of the 
     values of eccentricity and inclination observed for many Atiras \citep{2018RNAAS...2b..46D,2019MNRAS.487.2742D,2019RNAAS...3..106D}; 
     however, the value of the argument of perihelion does not librate but circulates. In striking contrast, the nominal orbit and one 
     control orbit that is most different from the nominal one show von~Zeipel--Lidov--Kozai oscillations for the future evolution of 
     2020~AV$_{2}$. This behaviour is associated with capture in the 3:2 mean-motion resonance with Venus at 0.552~au. For the nominal 
     orbit, capture into this resonance depends critically on close encounters with Mercury under 0.003~au in about 10 to 15~kyr from now. 
     For more distant encounters, this capture is not observed. We have found that control orbits arbitrarily close to the nominal one do 
     not lead to capture if close encounters with Mercury do not cross the critical value of nearly 0.003~au, which is still outside the 
     Hill radius of the planet. Although 2020~AV$_{2}$ is probably not yet trapped in resonance with Venus, it cannot be discarded that 
     other, not-yet-discovered Vatiras may be captured into the 3:2 mean-motion resonance with Venus, remaining there for extended periods 
     of time. Figure~~\ref{inv}, left-hand side set of panels, show that 2020~AV$_{2}$ may also become trapped inside the 29:1 mean-motion 
     resonance with Jupiter at 0.5507~au in the future and perhaps experienced trapping in the 29:12 mean-motion resonance with Earth at 
     0.5553~au in the past. The fact is that the Vatira orbital realm is very rich in closely spaced mean-motion resonances that may 
     contribute to keeping a relatively large population of minor bodies switching between the various resonant states.
%
%-------------------------------------------------------------------------------------------------------------------------------------------
%
%
%-------------------------------------------------------------------------------------------------------------------------------------------
%

  \section{Discussion}
     The calculations presented in the previous section indicate that the 3:2 mean-motion resonance with Venus may play an important role in
     the orbital evolution of the Vatira population. The dynamical context could be similar to that of the Hilda group asteroids that have 
     orbital periods 2/3 that of Jupiter. Hildas are numerous and include at least two collisional families \citep{2008MNRAS.390..715B}. As 
     with the Hildas and Jupiter, long-term stable members of the Vatira population with $a$$\sim$0.552~au may orbit the Sun reaching 
     aphelia opposite Venus, or 60{\degr} ahead or behind Venus. In the particular case of 2020~AV$_{2}$, it might be 60\degr ahead (or east 
     of) Venus. The 3:2 mean-motion resonance with Venus and its dynamical properties have been discussed within the context of ESA's Solar 
     Orbiter mission \citep{2007ESASP.641E...1M}.\footnote{\url{https://sci.esa.int/documents/34903/36699/1567259614917-MAO-WP-483_Solar-Orbiter-Mission-Analysis_i1r0_2005-11.pdf}} 
     \citet{2006Icar..184...29G}, top left-hand side panel of his fig.~7, has shown that the 3:2 mean-motion resonance with Venus is 
     relatively close in strength to that of the 1:1 with Venus, which is the strongest within the Atira orbital realm. 
%
%-------------------------------------------------------------------------------------------------------------------------------------------
%
     \begin{figure}
       \centering
        \includegraphics[width=\linewidth]{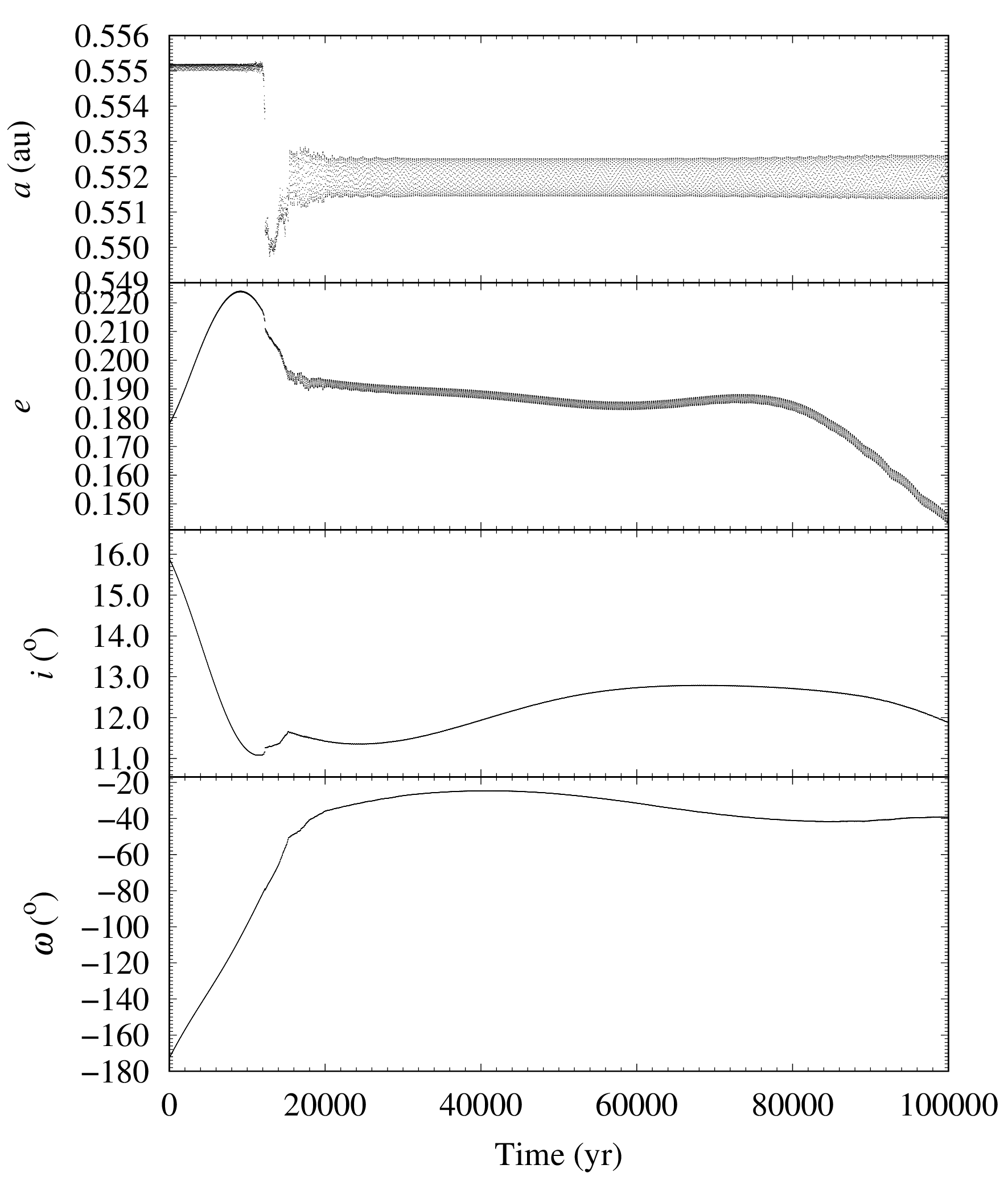}
        \caption{Similar to Fig.~\ref{inv} but for a different control orbit close to the $-$12$\sigma$ orbit. 
                }
        \label{invx}
     \end{figure}
%
%-------------------------------------------------------------------------------------------------------------------------------------------
%

\begin{figure}
       \centering
        \includegraphics[width=\linewidth]{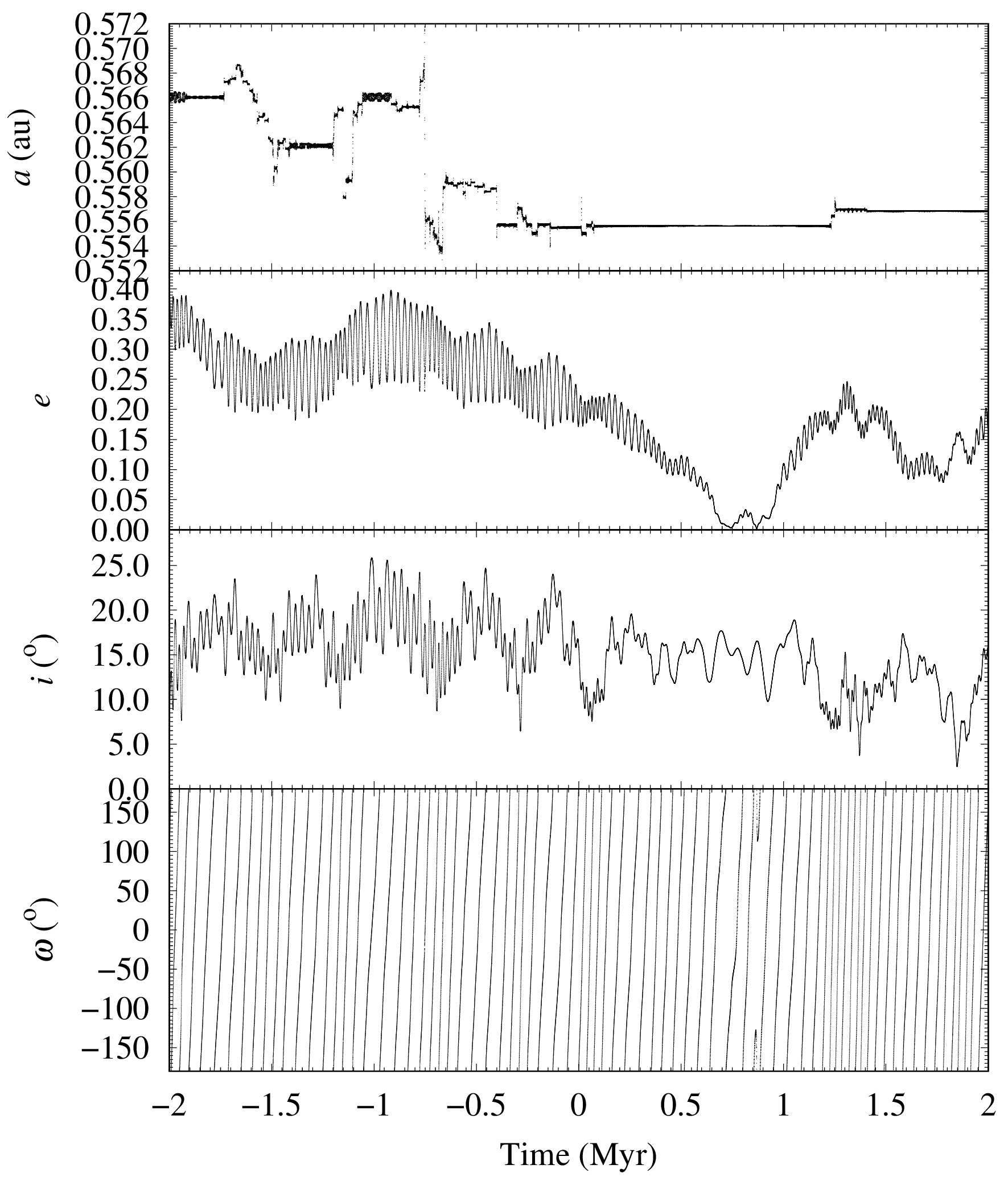}
        \caption{Similar to Fig.~\ref{inv} but for a different control orbit close to the $-$3$\sigma$ orbit and longer integration time. 
                }
        \label{invl}
     \end{figure}

     The onset of the von~Zeipel--Lidov--Kozai oscillations has been found for multiple instances on the control orbits. Figure~\ref{invx} 
     shows that even when the control orbit is not initially trapped into the 3:2 mean-motion resonance with Venus, there are viable 
     dynamical pathways that may naturally lead to it. If the resonance can be reached relatively easily, it may be well populated although 
     these objects have so far evaded detection because their discovery requires the observation at solar elongations below 45\degr. On the 
     other hand, the von~Zeipel--Lidov--Kozai oscillation may also be associated with other orbital configurations within the Vatira orbital 
     realm. Figure~\ref{invl} shows a longer calculation for a different control orbit where von~Zeipel--Lidov--Kozai oscillations are 
     observed at about 0.85~Myr into the future when the body was not trapped into the 3:2 mean-motion resonance with Venus but had a very 
     low value of the orbital eccentricity; in this case, the oscillation is about $\omega$=180{\degr} so the nodal points ---where the 
     orbit crosses the ecliptic--- are located at perihelion and at aphelion \citep{1989Icar...78..212M}.   
%
%-------------------------------------------------------------------------------------------------------------------------------------------
%
%
%-------------------------------------------------------------------------------------------------------------------------------------------
%

\begin{figure}
       \centering
        \includegraphics[width=\linewidth]{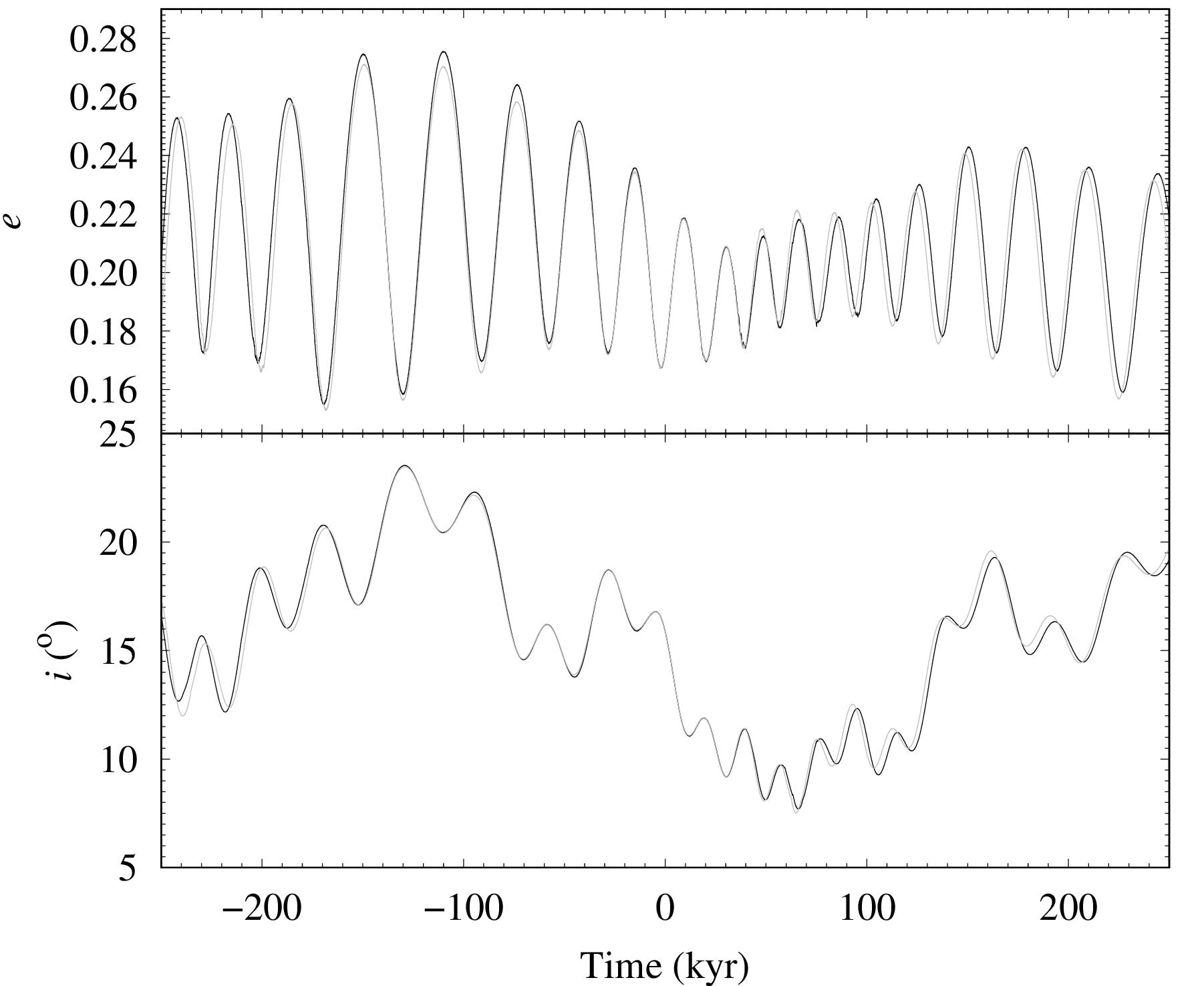}
        \caption{Evolution of the values of the eccentricity (top panel) and inclination (bottom panel) of one control orbit of 
                 2020~AV$_{2}$ arbitrarily close to the nominal one under the Newtonian (in black) and post-Newtonian (in grey) 
                 approximations.
                }
        \label{PN}
     \end{figure}

     \citet{2016MNRAS.458.4471R} argued that many Atiras remain on regular orbits for at least 1~Myr, Fig.~\ref{invl} shows that this may 
     also be the case for Vatiras. \citet{2016MNRAS.458.4471R} concluded that Atiras populate a very unstable region of the inner Solar 
     system that makes their dynamical evolution rather chaotic, Fig.~\ref{invl} shows that although their orbital evolution is unstable,
     with multiple changes in the value of $a$, each transition drives the minor body towards another orbit of the Vatira type. Although
     this changing evolution is consistent with chaos, the overall behaviour is significantly more quiet than that of typical near-Earth
     objects. As in the case of 2019~AQ$_{3}$ \citep{2019MNRAS.487.2742D}, Fig.~\ref{PN} shows that the overall evolution is preserved in 
     the post-Newtonian case. 
%
%-------------------------------------------------------------------------------------------------------------------------------------------
%
%
%-------------------------------------------------------------------------------------------------------------------------------------------
%

     On the other hand, a statistical analysis using the orbit model\footnote{\url{http://neo.ssa.esa.int/neo-population}} developed by the
     Near-Earth Object Population Observation Program (NEOPOP) and described by \citet{2018Icar..312..181G} shows that 2020~AV$_{2}$ is an 
     outlier (in terms of size) as discussed by \citet{2019MNRAS.487.2742D} for the case of 2019~AQ$_{3}$. It is unclear why so many Atiras
     (and now the first Vatira) have $H$$<$23--25~mag, see table~2 in \citet{2019MNRAS.487.2742D}.
  
  \section{Conclusions}
     In this Letter, we have carried out an exploration of the orbital evolution of 2020~AV$_{2}$, which is the first minor body ever found 
     following a path contained entirely within the orbit of Venus. This exploration has been performed using direct $N$-body simulations 
     under the Newtonian and post-Newtonian approximations and covers the orbital domain immediately adjacent to its latest orbit 
     determination. Although the orbit of 2020~AV$_{2}$ will certainly be improved in the future, our conclusions seem to be sufficiently 
     robust: 
     \begin{enumerate}[(i)]
        \item All the control orbits starting within 9$\sigma$ of its current nominal position and velocity show that it is a member of the 
              Vatira group and that it was probably part of the Atira dynamical class in the relatively recent past (about 10$^{5}$~yr ago). 
              It can experience relatively close encounters with Mercury that is its only present-day direct perturber.
        \item It is probably not under the influence of the von Zeipel-Lidov-Kozai mechanism, but it displays an anticoupled oscillation 
              of the values of eccentricity and inclination in such a way that when the eccentricity reaches its maximum, the inclination 
              reaches its minimum and vice versa.
        \item It is statistically possible that 2020~AV$_{2}$ could be currently subjected to the 3:2 mean-motion resonance with Venus, 
              although it is far more probable that this Vatira is currently just outside this resonance. 
        \item A viable dynamical pathway that can lead Vatiras into a 3:2 mean-motion resonance with Venus has been found. The capture into 
              a resonant state is the result of multiple encounters with Mercury. When trapped in the resonance, the von Zeipel-Lidov-Kozai 
              mechanism is active.
        \item Vatiras may also spend significant periods of time trapped in other nearby mean-motion resonances such as the 29:1 with 
              Jupiter at 0.5507~au or the 29:12 with Earth at 0.5553~au.
        \item The von Zeipel-Lidov-Kozai oscillation state can be reached via multiple dynamical pathways within Vatira orbital parameter 
              space not necessarily linked to the 3:2 mean-motion resonance with Venus.
     \end{enumerate}
     Our results indicate that the orbital dynamics of members of the Vatira class may be far richer than conventionally thought. Multiple 
     transitions between Vatira paths are possible for a minor body while still fully embedded in Vatira orbital space and this situation 
     may continue for millions of years. Each temporarily stable path is linked to mean-motion resonances with Venus, Earth or Jupiter.

     After this work was submitted to MNRAS Letters, a relevant paper was submitted by S. Greenstreet to astro-ph (astro-ph/2001.09083). 
     She arrives to similar conclusions regarding the orbital evolution of 2020~AV$_{2}$ discussed here, but using different techniques 
     \citep{2020MNRAS.493L.129G}.

  \section*{Acknowledgements}
     We thank the anonymous referee for her/his constructive report, S.~J. Aarseth for providing the code used in this research, Q. Ye for 
     comments on ZTF operations and other details of the discovery of 2020~AV$_{2}$, S. Greenstreet for comments on her own results on the
     orbital evolution of 2020~AV$_{2}$, O. Vaduvescu and M. Popescu for comments on the observational results obtained by EURONEAR for
     2020~AV$_{2}$, F. Roig, M.~N. De Pr{\'a}, A.~O. Ribeiro and S. Deen for comments on Atiras and Vatiras, and A.~I. G\'omez de Castro for 
     providing access to computing facilities. This work was partially supported by the Spanish `Ministerio de Econom\'{\i}a y 
     Competitividad' (MINECO) under grant ESP2017-87813-R. In preparation of this Letter, we made use of the NASA Astrophysics Data System 
     and the MPC data server.

  \bsp
  \label{lastpage}
\end{document}